\documentclass[12pt]{article}
\usepackage[cp866]{inputenc}
\usepackage[T2A]{fontenc}

\def\nqq{\hspace*{-2em}}

\def\Jl#1#2{{\it #1\/} {\bf #2},\ }

\def\NP#1 {\Jl{Nucl. Phys.}{#1}}
\def\PRD#1 {\Jl{Phys. Rev.}{D#1}}
\def\DAN#1 {\Jl{Dokl. AN SSSR}{#1}}
\def\GC#1 {\Jl{Grav. \& Cosmol.}{#1}}
\def\GRG#1 {\Jl{Gen. Rel. Grav.}{#1}}
\def\JETF#1 {\Jl{Zh. Eksp. Teor. Fiz.}{#1}}
\def\JMP#1 {\Jl{J. Math. Phys.}{#1}}
\def\PLA#1 {\Jl{Phys. Lett.}{A#1}}
\def\PLB#1 {\Jl{Phys. Lett.}{B#1}}
\def\PRL#1 {\Jl{Phys. Rev. Lett.}{#1}}

                         \def\bearr{\begin{eqnarray} \lal}
                       \def\ear{\end{eqnarray}}
\def\beq{\begin{equation}}             \def\earn{\nonumber \end{eqnarray}}
\def\eeq{\end{equation}}               \def\dst{\displaystyle}
\def\bear{\begin{eqnarray}}            \def\tst{\textstyle}
                         \def\lal{&&\nqq {}}

\def\eqdef{\stackrel{\rm def}=}        
\def\e{{\,\rm e}}                      
\def\d{\partial}                       \def\const{{\rm const}}
      \def\Half{{\dst\frac{1}{2}}}
      \def\half{{\tst\frac{1}{2}}}

\def\cns{conic space }                     
\def\nsm{non-linear sigma model }          
            \def\emt{energy momentum tensor }

\begin{document}

\begin{center}
\Large{\bf {Exact solutions of SO(3) non-linear sigma model in a
conic space background}}
\end{center}

\begin{center}
\bf {V. B. Bezerra and C. Romero}
\end{center}

\begin{center}
{\it \noindent
 Departamento de F\'isica, Universidade Federal da
Para\'iba, C.Postal 5008, Jo\~ao Pessoa, PB, 58059-970, Brazil
}  
\end{center}

\begin{center}
\bf{Sergey Chervon}
\end{center}

\begin{center}
{\it \noindent Department of Theoretical and Mathematical Physics,
Ulyanovsk State University, 42, Leo Tolstoy str., 432970,
Ulyanovsk, Russia}
\end{center}

\bigskip

\subsection*{Abstract}

 {\small We consider a nonlinear sigma model coupled to the metric
of a conic space. We obtain 
restrictions for a \nsm to be a source of the conic space.
We then  study \nsm in the conic space background. We
find coordinate transformations which reduce the chiral
fields equations in the conic space background to field equations
in Minkowski spacetime. This enables us to apply the same methods
for obtaining exact solutions in Minkowski spacetime
to the case of a conic spacetime. In the case the solutions depend on 
two spatial coordinates we employ Ivanov's geometrical ansatz.
We give a general analysis and also present classes of solutions in
which there is dependence on three and four coordinates.
We discuss with special attention the intermediate 
instanton and meron solutions and their analogous in the conic
space. We find differences in the total actions and
topological charges of these solutions and discuss the role of the
deficit angle.}
\bigskip

PACS-94: 04.20.-q \ 98.80.-Dr  

\noindent Key words: cosmic string, non-linear sigma model,
deficit angle, exact solutions


\section{Introduction}

In the late seventies and beginning of eighties much attention was 
devoted to \nsm formulated in two-dimensional Euclidian and
pseudo-Euclidian spaces\cite{belpol75,hpss76,pohlmeyer76,woo77,gross78}. 
The reason for that lies on the strong analogy with four-dimensional gauge theories 
(instanton solutions, asymptotic freedom, and so on (see, e.g
reviews \cite{perelomov87,wiegmann85,perelomov89})). In most cases non-linear
sigma models  have
been considered as a laboratory or a toy models with the following
applications of the developed methods and solutions to
the investigation of gauge theories. When we consider \nsm in spaces
with more than two dimensions the analogy with  gauge theories fails.
The fact that in four-dimensional case there exist no 
topologically nontrivial solutions led to various attempts to
extend the model to four dimensions via the 
introduction of additional interactions with a gauge or a 
gravitational field \cite{aff79,ghivis82}.

When \nsm coupled to gravitation were considered a new
method of obtaining exact solutions for a model of this type was
developed \cite{ivanov83tmf}. The method is based on a isometric
ansatz, which preserves the isometric motion in spacetime and
in the target space. In this way new exact solutions in
gravitational theory have been obtained
\cite{ivanov83tmf}\cite{che83iv}. This method  was then applied to
obtaining exact solutions in the case of a four-dimensional (4D) Minkowski spacetime,
as a connection between an isometric subgroup and a subgroup
of target space was postulated. Moreover when applied 
to two-dimensional (2D) \nsm the method gives a wide class of exact solutions
\cite{ivache87}, in which instanton and meron solutions arises  as
a special case of the subgroup of rotations.

Conic space, on the other hand, has first appeared in the context of a
cosmic string model\cite{vilen81}. It can be characterized by a space-time
metric with a Riemann-Christoffel curvature tensor which vanishes everywhere,
except in one point,where there is a  conical singularity. The local flatness
of this kind of space-time can induce several interesting effects. As  
examples of these effects we can mention gravitational lensing\cite{gott}, eletrostatic self-force
on an electric charge at rest\cite{linet} and the so-called gravitational 
Aharonov-Bohm effect\cite{ford} among others. 

In this paper we investigate the \nsm in the a space-time with conic-type singularity,
namely, the space-time of a cosmic string, and show how the obtained results differ 
from those obtained in flat Minkowski spacetime.

This paper is organized as follows. In section 2, we introduce the \nsm in a conical
geometry. In section 3, we consider the dynamic equations and discuss the symmetries 
and general properties of their solutions and introduce the Ivanov's 
geometrical ansatz. In section 4, we obtain the two-dimensional solutions. In section 5,
we discuss the four-dimensional solutions. Finally, in section 6, we draw some 
conclusions.

\section{Massive nonlinear sigma model coupled to the metric of the \cns }

It is well known that the spacetime sourced by a straight static
(cosmic) string, is a conical space with metric given by \cite{vilen81}
\beq
    ds^2=dt^2 -dz^2-d\rho^2 - (1-4G\mu)^2\rho^2d\phi^2
    =\varrho_{ik}dx^i dx^k \label{mcs}
\eeq
which possesses a conical singularity characterized by 
a deficit angle $\delta =8 \pi G\mu $. (We sometimes will use the notation $(1-4G\mu)^2=
\alpha^2$).

In this section we are going to consider a massive \nsm \cite{abraham91}, that
is, a \nsm with a potential of self-interaction, coupled to the
metric of the \cns (\ref{mcs}). As a target space we choose a two-dimensional 
sphere $S^2$ admitting SO(3)-symmetry, the metric of which is given by
\beq
ds^2=d\chi^2+ \sin ^2 \chi d\Theta^2=h_{AB}\varphi^A\varphi^B,
~~0<\chi < \pi ,~~0 \leq \Theta < 2\pi . \label{mts2}
\eeq
The dynamic equations of the chiral fields 
can be obtained form the variation of the action of \nsm with the potential
of self-interaction $W(\varphi^C)$

\beq
S_{NSM}=\int \sqrt{-g} d^4x \left( \,
 {1\over 2}\, h_{AB}\varphi_{,i}^A\varphi^B_{,k} \varrho^{ik}-W(\varphi^C)
 \right),\label{ac-nsm}
\eeq
with respect to the chiral fields $ \varphi^C $ and are given by 

\beq \frac{1}{\sqrt{-\varrho}}\partial_{\mu}\left(\sqrt{-\varrho}
\varrho^{\mu\nu}h_{AB}\varphi_{,\nu}^{B}\right)-
\frac{1}{2}\frac{\d h_{BC}} {\d \varphi ^{A}}
\varphi _{,\mu}^{B}\varphi _{,\nu}^{C}\varrho^{\mu\nu}+ \frac{\d
W}{\d \varphi^A} =0.\label{deq-nsm}
\eeq

On the other hand, the energy-momentum tensor of a massive \nsm 
can be calculated from the formula
\beq
T_{\mu\nu}=h_{AB}\varphi_{,\mu}^A\varphi_{,\nu}^B-\varrho_{\mu\nu}
\left(\half h_{AB}\varphi_{,\gamma}^A\varphi_{,\delta}^B
\varrho^{\gamma\delta} -W(\varphi^C)\right).\label{emt-nsm}
\eeq

If one considers the coupling of the chiral fields to the gravitational field,
then the Einstein equations 
\beq\label{einst3}
R_{\mu\nu}=\kappa \left( T_{\mu\nu}-\Half g_{\mu\nu}T \right).
\eeq
should be added. In our case we may set $g_{\mu\nu}=\varrho_{\mu\nu} $ in equation
(\ref{einst3}).

In the literature devoted to conic space (see, for example,
\cite{vilen81}) it has been stressed that cosmic strings have
very unusual properties. The fact, that self-gravitating \nsm also
may have very special solutions \cite{ch86a}, due to the
structure of \emt (\ref{emt-nsm}), gives hope to find possible
configurations of the target space of \nsm which lead to the \cns
(\ref{mcs}). For the sake of simplicity and with the aim to keep
effects coming from the deficit angle let us for a moment
reduce our consideration to the case when the chiral fields
$\chi$ and $\Theta$ are functions
of the coordinates $\rho$ and $\phi$ only. With this restriction the
non-vanishing components of the energy-momentum tensor are
\bear
T_{00}= \half\left(\chi_\rho^2+\sin^2 \chi \Theta_\rho^2+
\frac{1}{\alpha^2\rho^2}\{ \chi_\phi^2+\sin^2 \chi
\Theta_\phi^2\}\right) +W;  \label{t00}
\\
 T_{zz}= -\half\left(\chi_\rho^2+\sin^2 \chi \Theta_\rho^2+
\frac{1}{\alpha^2\rho^2}\{ \chi_\phi^2+\sin^2 \chi
\Theta_\phi^2\}\right) -W;
\label{tzz}
                                                                            \label{tzz}
\\
 T_{\rho\rho} = \half\left(\chi_\rho^2+\sin^2 \chi \Theta_\rho^2 -
 \frac{1}{\alpha^2\rho^2}\{ \chi_\phi^2+\sin^2 \chi \Theta_\phi^2\}
 \right)-W;                                                            \label{trr}
\\
  T_{\phi\phi} = \half\left(\chi_\phi^2+\sin^2 \chi \Theta_\phi^2 -
 \alpha^2\rho^2\{ \chi_\rho^2+\sin^2 \chi \Theta_\rho^2\}
   \right)-\alpha^2\rho^2 W;                                             \label{tpp}
\\
T_{\rho \phi} = \chi_\rho \chi_\phi + \sin^2 \chi
\Theta_\rho\Theta_\phi     \label{02}
                                                                           \label{trp}
\ear

To solve the Einstein equations (\ref{einst3}) in the \cns
(\ref{mcs}) with the \emt (\ref{t00})-(\ref{02}) we have to consider
 the strong restrictions:
\beq
W=0, ~~\chi = \const ,~\Theta = \const .\label{stres1}
\eeq
It is easy to see that only when (\ref{stres1}) holds all components of the \emt for
$SO(3)$-symmetric \nsm will be equal to zero, in accordance with
the vanishing of the Einstein tensor $G_{\mu\nu}$ 
in the \cns (\ref{mcs}). Nevertheless, the Einstein equations 
$G_{\mu\nu}=0$ can be satisfied without any reference to (\ref{stres1}),
if we change the signature of metric of the target space (\ref{mts2}) from Euclidian
to pseudo-Euclidian and put the following restrictions on the chiral fields
\beq
W=0,~~\chi^2_\rho=\sin^2 \chi \Theta_\rho^2,~~ \chi^2_\phi=\sin^2
\chi \Theta_\phi^2  \label{stres2}
\eeq

In this way the conic space (\ref{mcs}) can be sourced by \nsm
(\ref{ac-nsm}) with the constraints imposed by (\ref{stres2}) to the target
space. However, if we want to avoid these constraints we have to consider the
pure kinetic \nsm ($W=0$) in the background of the conic space, i.e., we 
have to work with the chiral field equations (\ref{deq-nsm}) without references 
to the Einstein equations (\ref{einst3}).

It should be mentioned here that finding solutions of \nsm
in the background of some spaces of GR is rather a complicated task.
In many cases the solutions can be found only in asymptotic form
or by numerical methods (for recent investigations see, for
example, \cite{aiclec97,lha00}).

\section{The dynamic equations of \nsm on the \cns background}

Taking into account the \cns and the target space metrics 
(\ref{mcs}) and (\ref{mts2}), respectively, one can write out the
dynamic equations (\ref{deq-nsm}) in terms of cylindrical
coordinates (\ref{mcs}). But it is possible considerably simplify the background field
equations (\ref{deq-nsm}), if we choose new variables
corresponding to the circle
cylindric coordinates (cc-coordinates) 
$$
u=\ln \rho ;~~-\infty <u<\infty; ~~ v=\alpha\phi ;~~ 0\leq v <
2\pi\alpha ;~~z=z;~~-\infty <z<\infty.
$$

In these coordinates the metric of the conic space (\ref{mcs})
 takes the form
\beq
ds^2=dt^2 -dz^2-e^{2u}\left( du^2 + dv^2 \right)= dt^2 -dz^2-ds^2_{cs}.
\eeq
Here we define the two-dimensional conic sector of the conic
space as $ ds^2_{cs} = e^{2u}\left( du^2 + dv^2 \right)$.

The dynamic equations (\ref{deq-nsm}) with $W=0$ can be written as

\bear
-\chi_{tt}+\chi_{zz}+\chi_{uu}+\chi_{vv}- \sin\chi\cos\chi~\left(
-\Theta_t^2+\Theta_z^2 + \Theta_u^2+\Theta_v^2 \right)=0;
\label{f1uv}
\\
-\Theta_{tt}+\Theta_{zz}+\Theta_{uu}+\Theta_{vv}+ 2\cot\chi \left(
-\chi_t\Theta_t+\chi_z\Theta_z +\chi_u\Theta_u+\chi_v\Theta_v
\right) =0. \label{f2uv}
\ear
Let us note that the equations (\ref{f1uv}-\ref{f2uv}) have the same form as the
equations of \nsm in the Minkowski space-time 
\cite{che85iv} because of conformal invariant symmetry of the
conic sector. Thus the solutions have the same form as in Minkwoski spacetime case
 but their properties will be rather different. In the next sections we will consider
 these solutions, obtained in \cite{ivache87} and
\cite{che85iv} for the case of $SO(3)$-symmetric \nsm in
Minkowski spacetime background.

\subsection{ Symmetries and general properties of the solutions}

First of all let us consider the symmetries of the chiral fields
equations (\ref{f1uv})-(\ref{f2uv}). These equations are invariant
with respect to the substitutions:
\beq
\chi \leftrightarrow -\chi ,~~\chi \leftrightarrow \chi+\pi k,~~k
\in Z;~~ \Theta \leftrightarrow -\Theta,~~\Theta
\leftrightarrow\Theta+\const ,~~\zeta\leftrightarrow\eta .
\label{sub}
\eeq
where the new variables $\zeta$ and $\eta $ are defined as
 $ \zeta = \half (z+t),~~\eta=\half (z-t).$
It is clear that the field equations (\ref{f1uv})-(\ref{f2uv})
are also invariant under the change 
 $u\leftrightarrow v\leftrightarrow z\leftrightarrow u.$
It is important to bear in mind that these substitutions can be always
done. Let us now turn our attention to the solutions which are related to the 
conic sector(u,v): $ds_{cs}^2=\e^{2u}\left( du^2 + dv^2\right)$ .

Let us start by mentioning some simple solutions and properties which
may be of interest as far as quantum properties are concerned:

i) Simple solutions corresponding to free fields, sometimes considered as classical
vacuum solutions in the two-dimensional case \cite{aff79}. These have
the same meaning as in the four-dimensional case and are given by:
$\chi=\chi_0=\const ,~~\Theta=\Theta_0=\const . $

ii) If the chiral fields $\chi$ and $\Theta$ depend on the two
coordinates $u$ and $v$ or $z$ and $t$, the dynamic equations correspond
to those in the two-dimensional Euclidian
\cite{belpol75} or pseudo-Euclidian \cite{gross78} \nsm ,
respectively. The solutions have the same form
 as in already mentioned works \cite{belpol75,gross78} for the
case of four-dimensional \nsm .

Let us include an additional dependence on $\zeta$ for the chiral
fields, namely  $\chi $ and $\Theta$ being  functions of
$u,v$ and $\zeta$. Then, it is clear from the equations
(\ref{f1uv})-(\ref{f2uv}), that any solution of two-dimensional \nsm
of the conic sector $(u,v)$ 
can be extended to the case with the dependence on $\zeta$. To
this end we can replace the constants of integration in two-dimensional 
 \nsm by arbitrary functions of $\zeta$.
%

\subsection{Ivanov's geometrical ansatz}

A procedure to obtain the exact solutions of the dynamic equations is based on
Ivanov's geometrical ansatz \cite{ivanov83tmf}. This method can be
described as follows. First of all, we have to postulate
Ivanov's geometrical ansatz. Let us suppose that 
the space-time admits the group of isometric and homothetic
symmetry $G_r$ with $r$ -linear independent vectors of isometric
or homothetic motion
$\xi_\alpha^i(x)$ and the structure's  constants 
$C_{\beta\gamma}^\alpha $. Let us remind that an infinitesimal
transformation $ \delta x^{i}=\xi^{i}(x)$ corresponds to an isometric
motion if $\xi^{i}(x)$ is the solution of the Killing equations
\beq \label{1-4.1.4} \xi_{\mu;\nu}+\xi_{\nu;\mu}=0. \eeq
where (;) denotes covariant derivative.

The vector field $\xi^{\mu}(x)$ defines a homothetic motion if
\beq \label{1-4.1.5}
\xi_{\mu;\nu}+\xi_{\nu;\mu}=\lambda g_{\mu\nu}, ~~\lambda=\const.
\eeq

We assume that the target space  admits a group of
isometric symmetry $G_s$ with $s$ -linear
independent Killing's vectors $\zeta_a^A(\varphi)$ and the structure's 
constants
$C_{b c}^a $.  Ivanov's geometrical ansatz consists in 
postulating the following relation between the symmetries of the basic and target
spaces
\beq  \label{1-4.1.6}
\xi_{\alpha}^{\mu}\partial_{\mu}\varphi^{A}=K^{a}_{\alpha}\zeta^A_a,
~K^{a}_{\alpha} =\const, \eeq
where
$$ \alpha=1,\ldots,r;~~a=1,\ldots,s.
$$

In the case under consideration, that is, when Einstein equations
are not being taken into account, the equation (\ref{1-4.1.6}) can be assumed
for some subgroups of the group $G_r$. It should be
noted that the constants $K^{a}_{\alpha}$ are not arbitrary. The
integrability conditions of (\ref{1-4.1.6}) implies that  
 $K^{a}_{\alpha}$ must satisfy:
$$
K^{a}_{\gamma}C^{\gamma}_{\alpha\beta}=K^{b}_{\alpha}K^{c}_{\beta}
C^{a}_{bc} .
$$

\noindent
By integrating the equations (\ref{1-4.1.4}) and
(\ref{1-4.1.5}) in Minkowski spacetime
\beq\label{mink}
dS^2=\eta_{\mu\nu}dx^\mu dx^\nu=-(dx^0)^2 + (dx^1)^2+ (dx^2)^2+
(dx^3)^2
\eeq we obtain the linear independent Killing vectors and
vectors of homothetic motion (conformal Killing vectors) in
Minkowski spacetime (\ref{mink}):

\beq \label{1-4.1.11}
 \xi_{1}^{\mu}=\delta_{1}^{\mu},~ \xi_{2}^{\mu}=\delta_{2}^{\mu},~
 \xi_{3}^{\mu}= x^{2}\delta_{1}^{\mu}-x^{1}\delta_{2}^{\mu},~
\xi_{4}^{\mu}=\frac{\lambda}{2}
 (x^{1}\delta_{1}^{\mu}+x^{2}\delta_{2}^{\mu})
\eeq

In the case of the conic space (\ref{mcs}) the Killing vectors and
conformal Killing vectors for conic sector are
\beq \label{kckvs}
\xi_{1}^{\mu}=\delta_{v}^{\mu},~ \xi_{2}^{\mu}=e^{-u}
 \left(\sin v \delta_{v}^{\mu}+\cos v \delta_{u}^{\mu}\right),~
\xi_{3}^{\mu}=e^{-u}
 \left(\sin v \delta_{v}^{\mu}-\cos v \delta_{u}^{\mu}\right),~
 \xi_{4}^{\mu}=\frac{\lambda}{2}\delta_{u}^{\mu}.
\eeq

It is clear that the vector $\xi_{1}^{\mu}$ corresponds to a
rotation and the vector $\xi_{2}^{\mu}$ and $ \xi_{3}^{\mu}$
corresponds to translations along the $x$-axis and $y$-axis,
respectively. The vector $\xi_{4}^{\mu}$ , on the other hand,
is the vector of homothetic motion.

Now, Killing vectors of the target space ${S}^{2}$ (\ref{mts2}) can
be represented in the form

\bear \label{1-4.1.12}
\zeta_{1}^{A}=-\sin\Theta\delta_{1}^{A}-\cos\Theta\cot\chi\delta_{2}^{A};\\
\nonumber
\zeta_{2}^{A}=\cos \Theta\delta_{1}^{A}-\sin \Theta\cot \chi\delta_{2}^{A}\\
\nonumber \zeta_{3}^{A}=\delta_{2}^{A}. \ear
Thus, the geometric ansatz  (\ref{1-4.1.6}) with the help of
(\ref{1-4.1.12}) will take the form:
\bear \label{1-4.1.13}
\xi_{\alpha}^{i}\chi_{i}=-K_{\alpha}^{1}\sin \Theta+K_{\alpha}^{2}\cos\Theta;\\
\nonumber \xi_{\alpha}^{i}\Theta_{i}=-K_{\alpha}^{1}\cos
\Theta\cot \chi-K_{\alpha}^{2} \sin \Theta\cot
\chi+K_{\alpha}^{3}. \ear

Our strategy to obtain solutions is the following: for
each Killing vector we will solve the ansatz (\ref{1-4.1.13})
and then will insert the obtained solutions into the fields
equations (\ref{f1uv})-(\ref{f2uv}). Integrating the equations
will lead to the exact solutions.

\section{Two-dimensional Solutions}

Let us start from the rotational symmetry of the conic sector, which
is described by the Killing vector $ \xi_1^i$ in (\ref{kckvs}). The
equations (\ref{1-4.1.13}) will take the form
\bear \label{1-4.1.14}
\chi_{v} =-a\sin \Theta +b\cos \Theta,\\
\nonumber \Theta_{v} =-\coth \chi \left(a\cos \Theta +b\sin \Theta
\right) + m.
\ear
Here we used the notation
 $ K_{1}^{1}=a $, $ K_{1}^{2}=b $, $ K_{1}^{3}=m $ .
In the case of other symmetries the derivatives in left hand side of
(\ref{1-4.1.14}) should be replaced by derivatives with respect to
the corresponding variables: $x=e^u \cos v,~y=e^u \sin v$ or $ u $.

To integrate the chiral field equations (\ref{f1uv})-(\ref{f2uv})
it is convenient to consider the special restrictions on the
values of the constants in the equations (\ref{1-4.1.14}). After
investigating the rotational symmetry in this
section we  will discuss the remaining symmetries: translations and
homothety with the same values of the constants in
(\ref{1-4.1.14}).

\subsection{Some particular solutions }

In this subsection we will consider solutions corresponding to some particular 
situations. 

\subsection*{A1:~$ a=b=m=0,~\xi_1^\mu=\delta_v^\mu .$}

By integrating (\ref{1-4.1.14}) one can find
$$\chi=\chi (u),~\Theta=\Theta (u). $$
Then, after integrating the equation (\ref{f2uv}) we obtain
the constraint
\beq
\sin^2 \chi\Theta_u = c_1,
\eeq
The remaining equation (\ref{f1uv}) in terms of cc-coordinates
becomes
\beq
\chi_{uu}-c_1^2 \frac{\cos\chi}{\sin^3\chi}=0.
\eeq
Thus, the solution of the chiral fields dynamic equations reads

\bear \label{a1}
\chi=\pm \arccos\left(\sqrt{\frac{c^2_2-c_1^2}{c_2^2}}\sin (|c_2|
u) \right)+ 2\pi k,~~ k\in Z,\\
\label{a2} \Theta=\arctan\left(\frac{c_1}{c_2} \tan(|c_2|u)
\right).
\ear

This solution, evidently, does not contain the angle deficit
$\delta $.

\subsection*{A2: $ a=b=m=0,~\xi_2^\mu =e^{-u}
(\sin v \delta_v^\mu+\cos v \delta_u^\mu). $}

The solution of the ansatz (\ref{1-4.1.14}) will be given by
$$
\chi=\chi(y),~\Theta=\Theta(y)
$$

Thus the solutions of the chiral field equations are given by
formula (\ref{a1})-(\ref{a2}) with the substitution %
 $u \rightarrow y=e^u \sin v$.

\subsection*{A3: $ a=b=m=0,~\xi_3^\mu =e^{-u}
(\sin v \delta_v^\mu-\cos v \delta_u^\mu). $}

The situation here is the same as in the subsection {\bf A2} with
the substitution $x \leftrightarrow y $.

\subsection*{A4: $ a=b=m=0,~\xi_4^\mu =\frac{\lambda}{2}\delta_u^\mu .$}

The case of the homothetic motion is similar to the case of the
rotation, that is, to the results of the subsection {\bf A1}. The
solution of the chiral fields dynamic equations is described by
the formulas (\ref{a1})-(\ref{a2}) with the substitution $v
\leftrightarrow u $. Note that this solution contains the deficit
angle $\delta $ because $v=\alpha \phi $ and $\alpha^2=
(1-\delta/2)^2 $.

\subsection*{B1: $ a=b=0;~ m\neq 0,~\xi_1^\mu=\delta_v^\mu .$}

By integrating (\ref{1-4.1.14}) with $m \ne 0 $ we obtain
\beq   \label{1-4.1.16} \chi =\chi (u), ~~\Theta =mv +f(u).
\eeq

From the substitution $ \eta = \cos\chi $
the chiral fields equation (\ref{f1uv}),(\ref{f2uv}) can be reduced to the
first-order differential equations 

\bear \label{1-4.1.17}
f_u = \frac{c_{0}}{1-\eta^{2}},\\
\nonumber \left( \eta_u \right)^2 =m^{2} \eta^{4} - \left(c_{1}
+c_{0}^{2} + 2m^{2}\right)\eta^{2} +c_{1} + m^{2},
\ear
The last equation admits a general solution in terms of elliptic
integrals of the first kind $\mathbf{F(k,\varphi)}$. The inverse
dependence for $u$ is given by the formula

\bear \label{ell-int}
u= \frac{A}{\sqrt{2}}\sqrt{\left(\frac{2\eta^2c}{A}+1\right)}
\sqrt{\left(\frac{2\eta^2c+B}{B-A}\right)}\sqrt{\left(\frac{2\eta^2c}{-A}\right)}\\
\nonumber
\mathbf{F}\Bigl[\sqrt{\left(\frac{2\eta^2c}{A}+1\right)
}, \sqrt{\left(\frac{A}{A-B}\right)}\Bigr]
\left(c\sqrt{c\eta^6+b\eta^4+a\eta^2}\right)^{-1}
\ear
where
\beq \label{ell-par}
A=b_1+\sqrt{b_1^2-4ac},~~B=b_1-\sqrt{b_1^2-4a_1c},
~~a_1=m^2+c_1,~~b_1=-(2m^2+c_0^2+c_1),~~c=m^2
\eeq
The deficit angle appears in the solution for $ \Theta $
given by (\ref{1-4.1.16}).

\subsection*{B2: $ a=b=0;~ m\neq 0~\xi_2^\mu =e^{-u}
(\sin v \delta_v^\mu+\cos v \delta_u^\mu).$}

Taking into account the replacement $\partial_v$ by $\partial_x$
one can write the solution of the ansatz (\ref{1-4.1.14}) in the
following way
$$
\chi=\chi(y),~\Theta=m x + f(y).
$$
Thus the solution of the chiral fields equations is given by
formulas (\ref{1-4.1.16})-(\ref{ell-par})
with the substitutions: %
 $u \rightarrow y=e^u \sin v, ~v \rightarrow x$.
This solution does not contain the deficit angle.

\subsection*{B3: $ a=b=0,~m\neq 0,~\xi_3^\mu =e^{-u}
(\sin v \delta_v^\mu-\cos v \delta_u^\mu). $}

The situation here is the same as in the subsection {\bf B2} if we
take into account the exchange $x \leftrightarrow y $.

\subsection*{B4: $ a=b=0, ~m\neq 0,~\xi_4^\mu =\frac{\lambda}{2}\delta_u^\mu .$}

By integrating (\ref{1-4.1.14}) with $m \ne 0 $ and taking into
account the replacement $\partial_v \rightarrow \partial_u$ we
obtain
\beq   \label{b4-ans}
\chi =\chi (v), ~~\Theta = \frac{2}{\lambda}m u +f(v).
\eeq

In this case the solution of the chiral fields dynamic equations
is given by the formulas(\ref{1-4.1.16})-(\ref{ell-par}) with the substitutions: %
$m \rightarrow \frac{2}{\lambda}m,~v \leftrightarrow u $. Note
that this solution contains the deficit angle $\delta $.

\subsection*{C1: $a^{2} + b^{2} \neq 0; ~ m=0;~\xi_1^\mu=\delta_v^\mu .$}

By integrating (\ref{1-4.1.14}) we obtain the result

 \bear \label{1-4.1.19}
\chi = \arccos \left[ \sin \mu (u) \sin \left( pv +
 B(u)\right) \right] ,\\
\nonumber \Theta =\theta_{0} + \arcsin \left( \frac{\sin \mu \cos
\{(pv + B)\}} {\sqrt{1-(\sin^2 \mu)\sin^2 (pv + B)}} \right), \ear
where $ p^{2} = a^{2} + b^{2} , \coth \theta_{0} =\frac{a}{b}. $

The chiral fields equations reduced to the equations of the functions $
\mu (u) $ and $ B(u)$ which will have the same form as
(\ref{1-4.1.17}) when we make the substitutions: $\eta \to \cos \mu , f
\to B , m^{2} \to p^{2}$. The solution (\ref{ell-int}) will also be
 valid with the substitutions above.

\subsection*{C2: $ a^{2} + b^{2} \neq 0; ~ m=0;~\xi_2^\mu =e^{-u}
(\sin v \delta_v^\mu+\cos v \delta_u^\mu).$}%

The solution of the ansatz (\ref{1-4.1.14}) is given by the
formula (\ref{1-4.1.19}) with the substitutions:
 $u \rightarrow y, ~v \rightarrow x$.
Thus the solution of the chiral fields equations is given by
formulas (\ref{1-4.1.19}), (\ref{1-4.1.17}),
(\ref{ell-int}),(\ref{ell-par}) with the substitutions: %
$u \rightarrow y, ~v \rightarrow x$.

\subsection*{C3: $a^{2} + b^{2} \neq 0; ~ m=0;~\xi_3^\mu =e^{-u}
(\sin v \delta_v^\mu-\cos v \delta_u^\mu). $}

The situation here is the same as in the subsection {\bf C2} with
the substitution $x \leftrightarrow y $.
\subsection*{C4: $ a^2+b^2 \neq 0;~ m=0;~\xi_4^\mu =\frac{\lambda}{2}\delta_u^\mu .$}

This case corresponds to the case presented in the paragraph {\bf
C1} with the multiplication of the constants $a,b,m $ by
$\frac{2}{\lambda}$ and the replacement $u\leftrightarrow v$ in
the solution. The deficit angle is also contained in this
solution.

\subsection*{D1: $ a^{2} + b^{2}\neq 0; ~m\neq 0, ~\xi_1^\mu=\delta_v^\mu .$}

This case corresponds to the most general solution of Ivanov
geometric ansatz (\ref{1-4.1.13}) for $SO(3)$-invariant \nsm . The
solution of (\ref{1-4.1.13}) is

\beq \label{1-4.1.20}
 k\left( u+B(v)\right) = \arctan
\left(\frac{s}{\pm \sqrt {1\mp A^{2}k^{2}(1+s^{2})}}\right)
-\arctan {s} , \eeq
where $ k^{2} = a^{2} + b^{2} + m^{2} ,~~ s =\frac{k}{m}\coth
{\omega } ,~~ \omega =\Theta + \theta_{0} ,~~ \tan \theta_{0} =
\frac{b}{a} ,~~ A^{2} = A^{2}(v) $.

For the minus sign under the square-root the following replacements
will be used
\beq \label{1-4.1.21} A^{2}k^{2} = {\sin^{2} {\alpha (v)}},~~
\gamma^{2} = m^{2} + p^{2}{\cos^{2} \alpha },~~ m^{2}\le
\gamma^{2} \le k^{2}. \eeq

For the plus sign under square-root, the replacements are

\beq \label{1-4.1.22} A^{2}k^{2} = {\sinh^{2} {\alpha (v)}},~~
\gamma^{2} = m^{2} + p^{2}{\cosh^{2} \alpha },~~ \nonumber
\gamma^{2} \ge k^{2}. \eeq
Then the field $\chi $ can be defined via $\omega $ from the
relation:
\beq  \label{1-4.1.23}
\chi =\arctan
\left(\frac{(m-\omega_{v}')}{p\sin \omega }\right). \eeq

Using the replacements (\ref{1-4.1.20})-(\ref{1-4.1.23}), 
the fields equations can be reduced to that of the first order
\bear \label{1-4.1.24}
B_{v}' = \frac{C_{1}}{1-\psi^{2}},\\
\nonumber (\psi_v)^2 = k^2 \psi^4-\left( k^{2} + k^{2}C_{1}^{2}
 + m^{2}C_{2}^{2}\right)\psi^2 + m^{2}C_{2}^{2} ,
\ear where $ \psi^{2} = \frac{m^{2}}{\gamma^{2}}. $

It is clear that with the help of the replacements
(\ref{1-4.1.20})-(\ref{1-4.1.23}) we have reduced the problem to the
system of equations (\ref{1-4.1.17}) which has been solved in the
subsection {\bf B1}. Actually, the solution of the second equation
in (\ref{1-4.1.24}) can be given by 
(\ref{ell-int}) with the substitutions
$$
a_1=m^2C_2^2,~~b_1=-\left( k^{2} + k^{2}C_{1}^{2}
 + m^{2}C_{2}^{2}\right),~~c=k^2.
$$

\subsection*{D2: $ a^{2} + b^{2} \neq 0; ~ m \neq 0;~\xi_2^\mu =e^{-u}
(\sin v \delta_v^\mu+\cos v \delta_u^\mu).$}%

In this case the solution of the chiral field equations
(\ref{f1uv})-(\ref{f2uv}) is given by 
(\ref{1-4.1.20})-(\ref{1-4.1.24}) with the substitutions:
 $u \rightarrow y, ~v \rightarrow x$.

\subsection*{D3: $a^{2} + b^{2} \neq 0; ~ m\neq 0;~\xi_3^\mu =e^{-u}
(\sin v \delta_v^\mu-\cos v \delta_u^\mu). $}

The situation here is the same as in the subsection {\bf D2} with
the substitution $x \leftrightarrow y $.

\subsection*{D4: $ a^2+b^2 \neq 0;~ m \neq 0;~\xi_4^\mu
=\frac{\lambda}{2}\delta_u^\mu .$}

This corresponds to the case presented in the subsection {\bf D1}
with the replacement $u\leftrightarrow v$. The solution is given
by formulas of the subsection {\bf D1} with the replacement
$u\leftrightarrow v$. The deficit angle also appears in this solution.

\subsection{Special solutions: the instanton and meron solutions}

Let us recall some general properties of the instanton and meron
solutions. Theses solutions are defined on 
two dimensional Euclidian space-time
$ dS^2_M = (dx^1)^2 + (dx^2)^2.$

The standard instanton solution is given as follows
\cite{ivache87}
\beq
\chi=2 \arctan\sqrt{(x^1)^2 + (x^2)^2}; ~~
\Theta=\arctan\frac{x^1}{x^2}.
\eeq

The well-known properties of the instanton solution are: the
finiteness of the action $S_{NSM}=4\pi$ and a topological charge
$Q$ which should is equal to unity ($Q=1$) \cite{perelomov87}. The
"topological charge density" can be defined as the Jacobian of the
mapping of the two-dimensional spacetime into the target space -- a two-dimensional
sphere in the case under consideration.

Let us consider the two-dimensional subspace (conic "plane") of
the \cns (\ref{mcs})
$$
ds_c^2=e^{2u}\{du^2+dv^2\}.
$$
For this case 
we can immediately obtain the analog of the instanton solution
\beq
\chi=2 \arctan e^u; ~~ \Theta= v.
\eeq

Simple calculations give for the action the value
$$
S=4\pi\alpha = S_{IS}-\delta,~~\delta= 8\pi G\mu.
$$
Here we denote by $S_{IS}$ the standard value of the action of
the instanton, which is equal to $4\pi$.

The topological charge corresponding to the conic "plane" will be given by
$$
Q=\alpha=\frac{2}{4\pi}(2\pi-\delta)=1-4G\mu =Q_{IS}-4G\mu.
$$

To obtain multi-instanton solutions we can set $v=\alpha
(\phi+2\pi k), k \in Z$. Clearly, in this case the deficit angle accumulates.

The meron solution is a solution which has a topological charge 
$\Half$ after regularization \cite{aff79}. The action
of the meron solution is logarithmically divergent. In our
case the meron solution is represented as being given by 
\beq \label{meron}
\chi =\frac{\pi}{2}, ~~ \Theta = \arctan \left(
\frac{u}{v}\right).
\eeq

It is possible to check that the action of the meron solution in
the conic "plane" (\ref{meron}) will be logarithmically divergent,
but the topological charge, through a regularization procedure in the
external three dimensional space, will be equal to $\Half\alpha $.
Thus the topological charge will also have an accumulative effect.

Multi-instanton and multi-meron solutions can be derived from
of the usual procedure for complex variable in external three
dimensional space. It is clear that these solutions will accumulate
a deficit angle effect in the invariant characteristics when the
integration over the angle coordinate $v$ is carried out.

\section{Four-dimensional solutions}

The geometric ansatz method described above has been applied to 
obtain exact solutions  in the case of SO(3) \nsm
in Minkowski space-time\cite{che85iv}. We will use the results obtained  in
\cite{che85iv} to the case of  \nsm in  the conic space background.

\subsection{Solutions for one-parametric subgroups}

The solutions of this kind can be represented in the form
\beq
\chi=\chi (\xi),~~\xi=a_\mu x^\mu,~~\Theta=\Theta (c_*
\xi),~~a_\mu a^\mu=0,~~c_* =\const ,
\eeq

where here $x^i$ denote:
Cartesian coordinates $x,y$, polar
coordinates $\rho, \phi$ or the coordinates $u=\ln \rho , \phi $. These
solutions can be obtained, of course, only when we have
pseudo-Euclidian signature of space-time; thus, the existence of an
isotropic vector $\vec{a}$ is necessary.

\subsection{The two-parametric subgroups solutions}

The system of equations (\ref{f1uv})-(\ref{f2uv}) coincides with
the system correspondign to SO(3) \nsm , considered in the background of 
Minkowski space-time. The solution, obtained for the last case,
can be applied in our case, which in cc-coordinates reads
\beq\label{gensol}
\chi=\arccos \left(\cos a \cos F(t,z,u,v)\right);~~\Theta =
\arctan\left(\tan F(t,z,u,v) / \sin a \right),
\eeq
where $a$ is a parameter and the function $F(t,z,u,v)$ must be a 
solution of the four-dimensional d'Alembert equation, that is,  the function $F$ must
satisfy the equation
\beq\label{L-F}
\triangle F(t,z,u,v)\eqdef F_{tt}-F_{zz}- F_{uu}-F_{vv}=0.
\eeq

As an example of a two-dimensional solution we can take the special
solution of Laplace equation (\ref{L-F}) in the form
\beq\label{solFuv}
F(u,v)=\Half \ln (u^2+v^2) +\arctan(u/v).
\eeq
It is clear that the deficit angle appears in this solution.

\subsection{Three-parametric static solution}

According to our presentation of the chiral field equations
(\ref{deq-nsm}) the three-parametric solution \cite{che85iv} takes
the form
\bear
\chi=\arccos\left(\frac{z}{\sqrt{u^2+v^2+z^2}} \right) \\
\Theta=\arccos\left( \frac{u}{\sqrt{u^2+v^2}}\right)
\ear

Because of the symmetry with respect to the coordinates $u$ and $v$
we can obtain the analog of the solution above in the form
\bear
\chi=\arccos\left(\frac{z}{\sqrt{v^2+u^2+z^2}} \right) \\
\Theta=\arccos\left( \frac{v}{\sqrt{v^2+u^2}}\right)
\ear

\section{Conclusions}

In spite of the close analogy between the solutions of a \nsm dynamic
equations in the case of Minkowski space-time and in the conic space, they
are, however, essentially different in the case of the instanton and meron
solutions. Perhaps these differences should be further investigated given
 the important role these solutions  have played  in the developing of the
quantum theory of \nsm .

\section*{Acknowledgements}

S.Chervon thanks CAPES for financial support during his recent visit to 
the Federal University of Paraiba. V. B. Bezerra and C. Romero would like to 
acknowledge CNPq and FAPESQ-Pb/CNPq(PRONEX) for partial financial support.

\end{document}